\documentclass[aps,prd,twocolumn,preprintnumbers,amsmath,amssymb,superscriptaddress,floatfix,showpacs,10pt]{revtex4-1}

\usepackage{color}
\usepackage[hyperindex,breaklinks]{hyperref}
\usepackage{graphicx}

\begin{document}

\preprint{DESY 13-068}

\title{Evolution of Helical Cosmic Magnetic Fields as Predicted by Magnetohydrodynamic Closure Theory}

\author{Andrey Saveliev}
\email{andrey.saveliev@desy.de}
\affiliation{{II}. Institut f\"ur Theoretische Physik, Universit\"at Hamburg, Luruper Chaussee 149, 22761 Hamburg, Germany}
\author{Karsten Jedamzik}
\email{karsten.jedamzik@um2.fr}
\affiliation{Laboratoire Univers et Particules de Montpellier, UMR5299-CNRS, Universit\'e Montpellier II, 34095 Montpellier, France}
\author{G\"unter Sigl}
\email{guenter.sigl@desy.de}
\affiliation{{II}. Institut f\"ur Theoretische Physik, Universit\"at Hamburg, Luruper Chaussee 149, 22761 Hamburg, Germany}

\begin{abstract}
We extend our recent derivation of the time evolution equations for the energy content of magnetic fields and turbulent motions for incompressible, 
homogeneous, and isotropic turbulence to include the case of nonvanishing helicity. These equations are subsequently numerically integrated in order 
to predict the present day primordial magnetic field strength and correlation length, depending on its initial helicity and magnetic energy density. 
We find that all prior analytic predictions for helical magnetic fields, such as the epoch when they become maximally helical and their subsequent 
growth of correlation length $L\sim a^{1/3}$ and decrease of magnetic field strength $B\sim a^{-1/3}$ with scale factor $a$, are well confirmed by the 
simulations. An initially fully helical primordial magnetic field is a factor $4\times 10^4$ stronger at the present epoch then its nonhelical 
counterpart when generated during the electroweak epoch.
\end{abstract}

\doi{10.1103/PhysRevD.87.123001}

\maketitle

\section{Introduction}
The time evolution of primordial magnetic fields from very early times to the present is a subject of continuing research (for reviews see 
\cite{PhysRep.348.163,Subramanian:2009fu,PhysRevD.80.123012,DuNe}). It has become of particular importance after the recent observational claim 
\cite{Neronov02042010} that a large fraction of the Universe may be filled by a magnetic field. A volume-filling cosmic magnetic field is a natural 
prediction when magnetogenesis occurred in the early Universe. It is thus of interest to have as precise as possible predictions on the evolution and 
possible magnitude of such fields.

There has already been much effort put into this investigation \cite{Dimopoulos:1996nq,Brandenburg:1996fc,Jedamzik:1996wp, Subramanian:1997gi,Son:1998my,
Banerjee:2004df,Campanelli:2007tc,Kahniashvili:2010gp}. However, there is an important fundamental difficulty which needs to be overcome; in particular, 
current numerical simulations do not reach the resolution required in order to reliably study magnetic fields over the large range of length scales 
and immense cosmic expansion factor between the early epoch of magnetogenesis and the recombination or present epoch. After all, the Universe changes 
in size by a factor of $10^{12}$ between the electroweak and the recombination epoch, while the respective dynamical time scales differ by  a factor of 
$10^{24}$. Only a small fraction of the initial magnetic energy survives dissipation - the exact amount may depend on the large scale tail 
of the magnetic field spectrum as has been seen to be the case for nonhelical fields. Such tails are extremely difficult to deduce from numerical 
simulations.

Though fairly detailed analytic estimates for the evolution of primordial magnetic fields exist \cite{Son:1998my,Banerjee:2004df,Campanelli:2007tc}, 
it may be of interest to verify them by more sophisticated methods. One such method is a closure theory of magnetohydrodynamics, which has been recently 
applied by us \cite{Saveliev:2012ea} to nonhelical magnetic fields, with the important result that the field strength tail dependence on the scale $L$ of 
$B(L)\sim L^{-\alpha/2}$ where $\alpha = 5$ (cf.~also \cite{Durrer:2003ja}) seems to be essentially independent of the initial conditions in incompressible 
magnetohydrodynamics (MHD) where we assume freely decaying turbulence after the magnetogenesis epoch, which, however, is dominated by decay times at least 
of the order of the Hubble time. Here the tail is important as it has a large influence over how much magnetic energy survives.

In this paper we extend the formalism developed in \cite{Saveliev:2012ea} to include helicity. Helical magnetic fields have already attracted much 
prior attention \cite{CambridgeJournals:373402,Field:1998hi,Christensson:2000sp,1475-7516-2009-11-001,Sigl:2002kt,0004-637X-640-1-335,Durrer:2003ja} 
as they may be much stronger due to an inverse cascade.

To do so this paper is structured as follows: in Sec.~\ref{sec:TimeHomoIsoHelMHD} we give a short derivation of the time-evolution equations for a 
homogeneous and isotropic medium including all terms which appear due to a nonvanishing magnetic helicity and apply the findings to the situation in 
the early Universe. More details can be found in Appendix \ref{app:calc}. In Sec.~\ref{sec:Results} we then present our results of the time evolution 
of primordial magnetic fields according to our equations for different initial conditions and finally draw our conclusions in Sec.~\ref{sec:Conclusions}.

\section{Time Evolution of the Magnetic and Kinetic Energy Content in Homogeneous Isotropic Magnetohydrodynamics} \label{sec:TimeHomoIsoHelMHD}
For an incompressible fluid (i.e., $\nabla\cdot\mathbf{v}=0$) the equations for the time evolution of the two main observables of magnetohydrodynamics, 
the velocity field of the turbulent medium $\mathbf{v}$ and the magnetic field $\mathbf{B}$, are given by
\begin{equation} \label{DiffB}
\partial_{t} \mathbf{B} = \frac{1}{4 \pi \sigma} \Delta \mathbf{B} + \nabla \times \left( \mathbf{v} \times \mathbf{B} \right) 
\end{equation}
and
\begin{equation} \label{Diffv}
\partial_{t} \mathbf{v} = - \left( \mathbf{v} \cdot \nabla \right) \mathbf{v} + \frac{\left( \nabla \times \mathbf{B} \right) \times \mathbf{B}}{4 \pi \rho} + \mathbf{f}_{v}\,,
\end{equation}
respectively. Here $\sigma$ is the conductivity (which is assumed to be very large in the following) and $\rho$ the mass density of the fluid, while 
$\mathbf{f}_{v}$ is some viscous density force. Note that for most part of the evolution of the early Universe, incompressibility is an excellent 
assumption due to the large speed of sound in the relativistic plasma.

\begin{figure*}
\centering
  \includegraphics[scale=0.467]{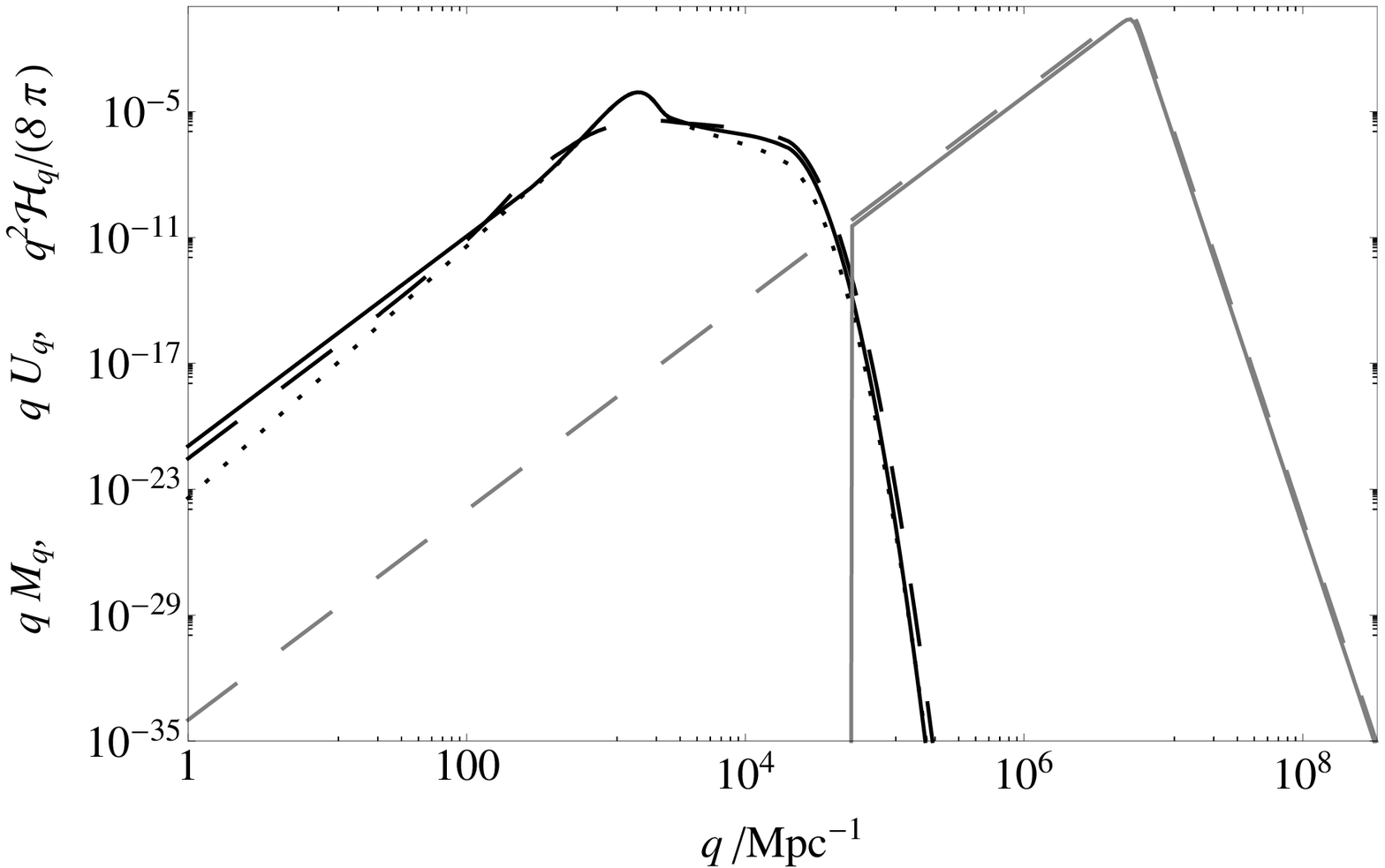}
  \includegraphics[scale=0.467]{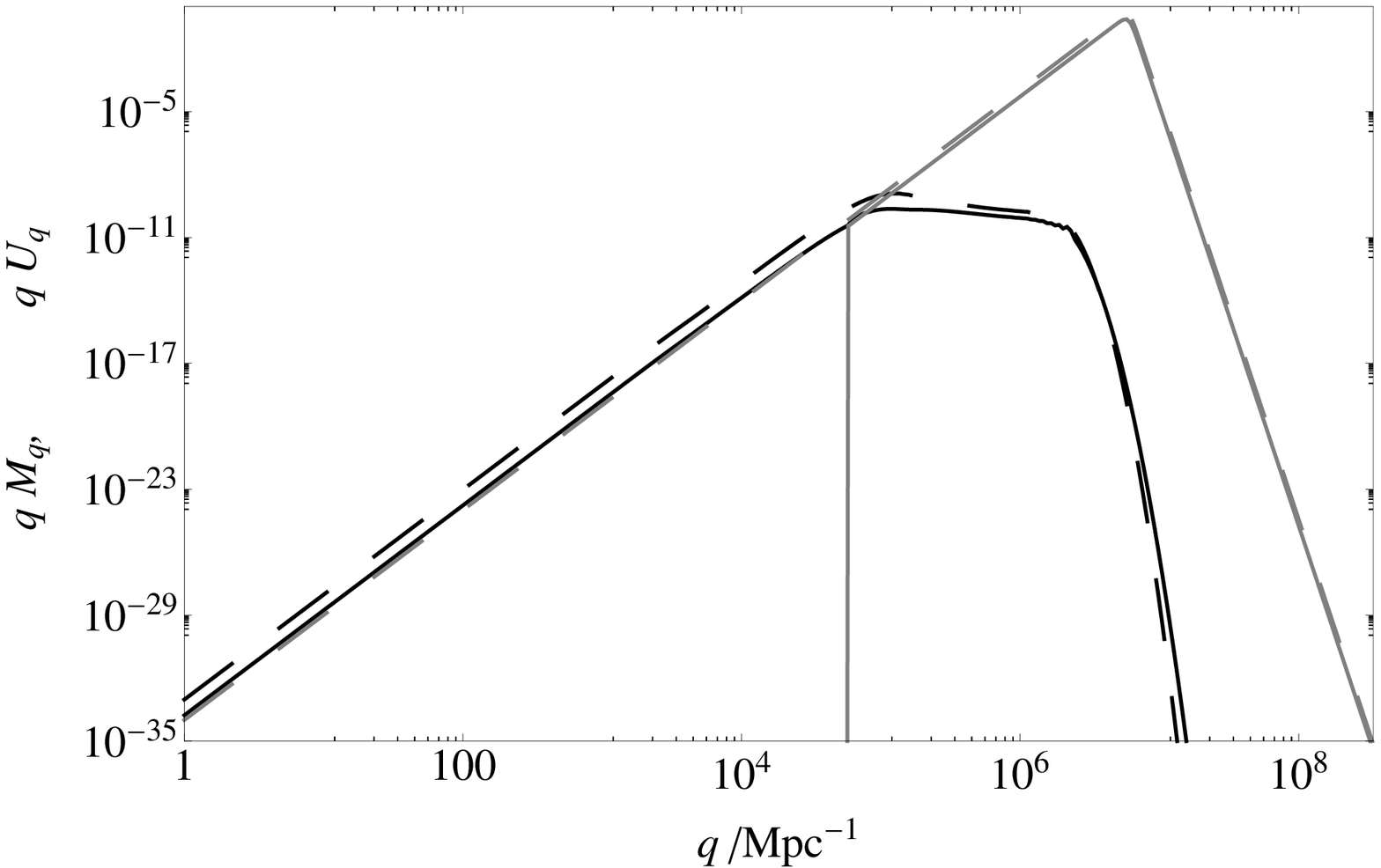}
  \caption{Time evolution of magnetic (solid) and kinetic (dashed) spectral energies, as well as of the spectral magnetic helicity (dotted). Gray lines 
           denote the initial conditions (i.e., at $a=1$) while the black lines represent the situation for $a=10^{6}$. In the left panel maximal 
           helicity has been assumed, while in the right panel the case for vanishing initial helicity is shown for comparison.}
  \label{fig:timeevolution}
\end{figure*}

Our main interest lies in the average buildup of magnetic and kinetic energy density, respectively, as well as of the magnetic helicity taken over an
ensemble of cosmic realizations (denoted by chevrons $\left\langle \right\rangle$), i.e.,
\begin{equation} \label{dtMdtU}
\left\langle \partial_{t} M_{q} \right\rangle,~\left\langle \partial_{t} U_{q} \right\rangle,~\left\langle \partial_{t} \mathcal{H}_{q} \right\rangle
\end{equation}
with $M_{q}$ being the magnetic spectral energy density defined through
\begin{equation} \label{MagEn}
\epsilon_{B} = \frac{1}{8 \pi}\int \frac{{\rm d^{3}}x}{V}\, \mathbf{B}^{2}(\mathbf{x}) = \int \frac{{\rm d^{3}}k}{8\pi}\, |\hat{\mathbf{B}}(\mathbf{k})|^{2} \equiv \rho \int {\rm d}k\, M_{k}\, ,
\end{equation}
$U_{q}$ the kinetic spectral energy density given by
\begin{equation} \label{KinEn}
\epsilon_{K} = \frac{\rho}{2} \int \frac{{\rm d^{3}}x}{V}\, \mathbf{v}^{2}(\mathbf{x}) = \frac{\rho}{2} \int {\rm d^{3}}k\, |\hat{\mathbf{v}}(\mathbf{k})|^{2} \equiv \rho \int {\rm d}k\, U_{k}
\end{equation}
and $\mathcal{H}_{q}$ the magnetic helicity density, i.e., 
\begin{equation}
\begin{split}
h_{B} &= \int \frac{{\rm d^{3}}x}{V}\, \mathbf{A}(\mathbf{x}) \cdot \left( \nabla \times \mathbf{A}(\mathbf{x}) \right) = \int \frac{{\rm d^{3}}x}{V}\, \mathbf{A}(\mathbf{x}) \cdot \mathbf{B}(\mathbf{x}) \\
&= i \int {\rm d^{3}} k \left\{ \left(\frac{\mathbf{k}}{k^{2}} \times \hat{\mathbf{B}}(\mathbf{k}) \right) \cdot \hat{\mathbf{B}}(\mathbf{k})^{*} \right\} \equiv \rho \int {\rm d}k\, \mathcal{H}_{k}
\end{split}
\end{equation}
where $\mathbf{A}$ is the magnetic vector potential and $h_{B}$ is the total magnetic helicity density, and where $\nabla\cdot\mathbf{A}=0$, i.e.,
Coulomb gauge, has been assumed.

For all expressions, $V$ denotes the volume and we have assumed cosmic homogeneity and isotropy which implies that $M_{k}$, $U_{k}$ and $\mathcal{H}_{k}$ 
are functions only of the magnitude $k$ of the wave vector $\mathbf{k}$. Furthermore, Parseval's Theorem has been used in all three cases in order to 
obtain a $k$ integral where a hat denotes the Fourier Transform normalized by $V^{\frac{1}{2}}$ (cf.~Appendix A of Ref.~\cite{Saveliev:2012ea}). With 
these assumptions we find
\begin{eqnarray} 
\label{MUK1}
&M_{q} = \frac{q^{2}}{2 \rho} |\hat{\mathbf{B}}(\mathbf{q})|^{2}\,, \\
\label{MUK2}
&U_{q} = 2 \pi q^{2} |\hat{\mathbf{v}}(\mathbf{q})|^{2}\,, \\
\label{MUK3}
&\mathcal{H}_{q} = \frac{4 \pi i}{\rho} \left(\mathbf{q} \times \hat{\mathbf{B}}(\mathbf{q}) \right) \cdot \hat{\mathbf{B}}(\mathbf{q})^{*}\,.
\end{eqnarray}

\begin{widetext}
By performing the calculations which are presented in Ref.~\cite{Saveliev:2012ea} and Appendix \ref{app:calc} of this paper, we obtain a very general 
result for the homogeneous and isotropic case, which for (\ref{dtMdtU}) is given by
\begin{equation} \label{dtMqH}
\begin{split} 
\left\langle \partial_{t} M_{q} \right\rangle = \int_{0}^{\infty} {\rm d}k \Bigg( &\Delta t \Bigg\{ -\frac{2}{3} q^2 \left\langle M_{q} \right\rangle \left\langle U_{k} \right\rangle - \frac{4}{3} q^{2} \left\langle M_{q} \right\rangle \left\langle M_{k} \right\rangle + \frac{1}{3} \frac{1}{(4 \pi)^{2}} q^{2} k^{2} \left\langle \mathcal{H}_{q} \right\rangle \left\langle \mathcal{H}_{k} \right\rangle \\
&+ \int_{0}^{\pi} { \rm d}\theta \left[\frac{1}{2} \frac{q^{4}}{k_{1}^{4}} \left( q^{2} + k^{2} - q k \cos\theta \right) \sin^{3}\theta \left\langle M_{k}\right\rangle \left\langle U_{k_{1}}\right\rangle \right] \Bigg\} \Bigg)
\end{split}
\end{equation}
and
\begin{equation} \label{dtUqH}
\begin{split}
\left\langle \partial_{t} U_{q} \right\rangle &= \int_{0}^{\infty} {\rm d}k \Bigg(\Delta t \Bigg\{ - \frac{2}{3} q^{2} \left\langle M_{k} \right\rangle \left\langle U_{q} \right\rangle - \frac{2}{3} q^{2} \left\langle U_{q} \right\rangle \left\langle U_{k} \right\rangle + \int_{0}^{\pi} {\rm d}\theta \Bigg[ \frac{1}{4} \frac{q^{3} k}{k_{1}^{4}} \left( q k \sin^{2}\theta + 2 k_{1}^{2} \cos\theta \right) \sin\theta \left\langle M_{k} \right\rangle \left\langle M_{k_{1}} \right\rangle \\
&+ \frac{1}{4} \frac{q^{4} k}{k_{1}^{4}} \left( 3 k - q \cos\theta \right) \sin^{3}\theta \left\langle U_{k} \right\rangle \left\langle U_{k_{1}} \right\rangle + \frac{1}{(16 \pi)^{2}} \frac{q^{3} k^{2}}{k_{1}^{2}} \left( - 2 q - q \sin^{2}\theta + 2 k \cos\theta \right) \sin\theta \left\langle \mathcal{H}_{k} \right\rangle \left\langle \mathcal{H}_{k_{1}} \right\rangle \Bigg] \Bigg\} \Bigg)
\end{split}
\end{equation}
as well as
\begin{equation} \label{dtHqH}
\langle \partial_{t} \mathcal{H}_{q} \rangle = \int_{0}^{\infty} {\rm d}k \Bigg(\Delta t \Bigg\{ \frac{4}{3} k^{2} \langle M_{q} \rangle \langle \mathcal{H}_{k} \rangle - \frac{4}{3} q^{2} \langle M_{k} \rangle \langle \mathcal{H}_{q} \rangle - \frac{2}{3} q^{2} \langle U_{k} \rangle \langle \mathcal{H}_{q} \rangle + \int_{0}^{\pi} { \rm d}\theta \Bigg[\frac{1}{2} \frac{q^{4} k^{2}}{k_{1}^{4}} \sin^{3}\theta \left\langle U_{k_{1}} \right\rangle \left\langle \mathcal{H}_{k} \right\rangle \Bigg] \Bigg\} \Bigg)\,,
\end{equation}
\end{widetext}
where $q$, $k$, and $k_{1}$ are the magnitudes of the wave vectors $\mathbf{q}$, $\mathbf{k}$ and $\mathbf{k_{1}} = \mathbf{q} - \mathbf{k}$, respectively, 
and $\theta$ is the angle between $\mathbf{q}$ and $\mathbf{k}$, i.e., $\mathbf{q} \cdot \mathbf{k} = q\,k\,{\rm cos\,\theta}$. Equations (\ref{dtMqH})-(\ref{dtHqH}) 
are a set of well-defined equations, since they ensure conservation of energy, momentum, mass, and helicity density to the lowest nontrivial order in
$\Delta t$.

It is important to note here that whereas the ideal MHD equations conserve energy and helicity, in any real application to astrophysics and cosmology
dissipation plays an important role. In particular, the early Universe is described by a large Prandtl number~\cite{Banerjee:2004df}, a measure of the 
relative importance of kinetic dissipation versus magnetic diffusion. Dissipation is thus dominated by shear viscosity, such that magnetic diffusion 
can be neglected. Since energy dissipates by either kinetic or magnetic dissipation, but helicity only by magnetic dissipation, in the early Universe 
helicity is essentially conserved whereas energy is not. To model this in our simulations we introduce a viscosity term into Eq.~(\ref{dtUqH}).

In addition it should be pointed out that the result obtained in \cite{Saveliev:2012ea} is consistent with the equations presented here as it can be 
seen that for vanishing initial magnetic helicity, it will be zero for all times and therefore simply the third term in (\ref{dtMqH}) and the last term 
in (\ref{dtUqH}) have to be dropped.

In order to study the time evolution of magnetic fields in the early Universe, we include expansion accounted by the scale factor $a$ with $a_0=1$ at 
the initial magnetogensis epoch. In Section III of Ref.~\cite{Saveliev:2012ea}, following also Ref.~\cite{Banerjee:2004df}, it was shown that 
Eqs.~(\ref{dtMqH})-(\ref{dtHqH}) remain invariant in the expanding Universe when time derivatives $\partial/\partial t$ are 
replaced by scale factor derivatives $(H_0/a)\,\partial/\partial \ln a$ and physical quantities in the integrands of Eqs.~(\ref{dtMqH})-(\ref{dtHqH}) 
are replaced by their comoving analogues. These comoving quantities (marked by a ``$c$'') are related to the physical ones by 
${\rm d}t_{c}={\rm d}t\, a^{-1}$, $v_{c}=v$, $\rho_{c}=\rho\,a^4$, $k_{c}=k a$, $M_{q}^{c} = M_{q} a^{-1}$, $U_{q}^{c}=U_{q} a^{-1}$, and $\mathcal{H}_{q}^{c}=\mathcal{H}_{q} a^{-2}$, 
whereas $H_0$ is the Hubble constant at the initial epoch. In the following the index ``$c$'' is dropped, and all quantities are meant to be the comoving 
ones unless noted otherwise.

\begin{figure*}
\centering
  \includegraphics[scale=0.467]{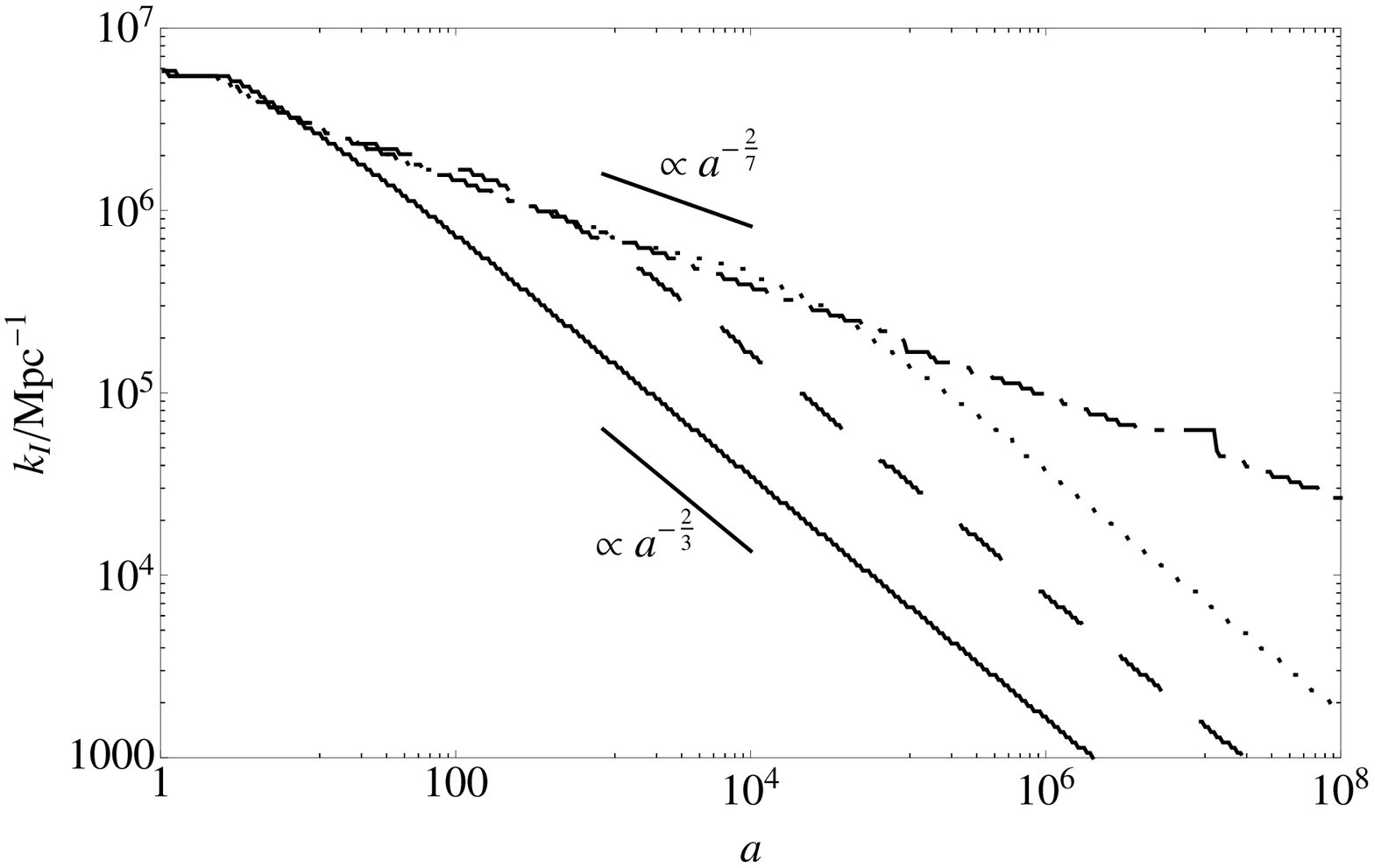}
  \includegraphics[scale=0.467]{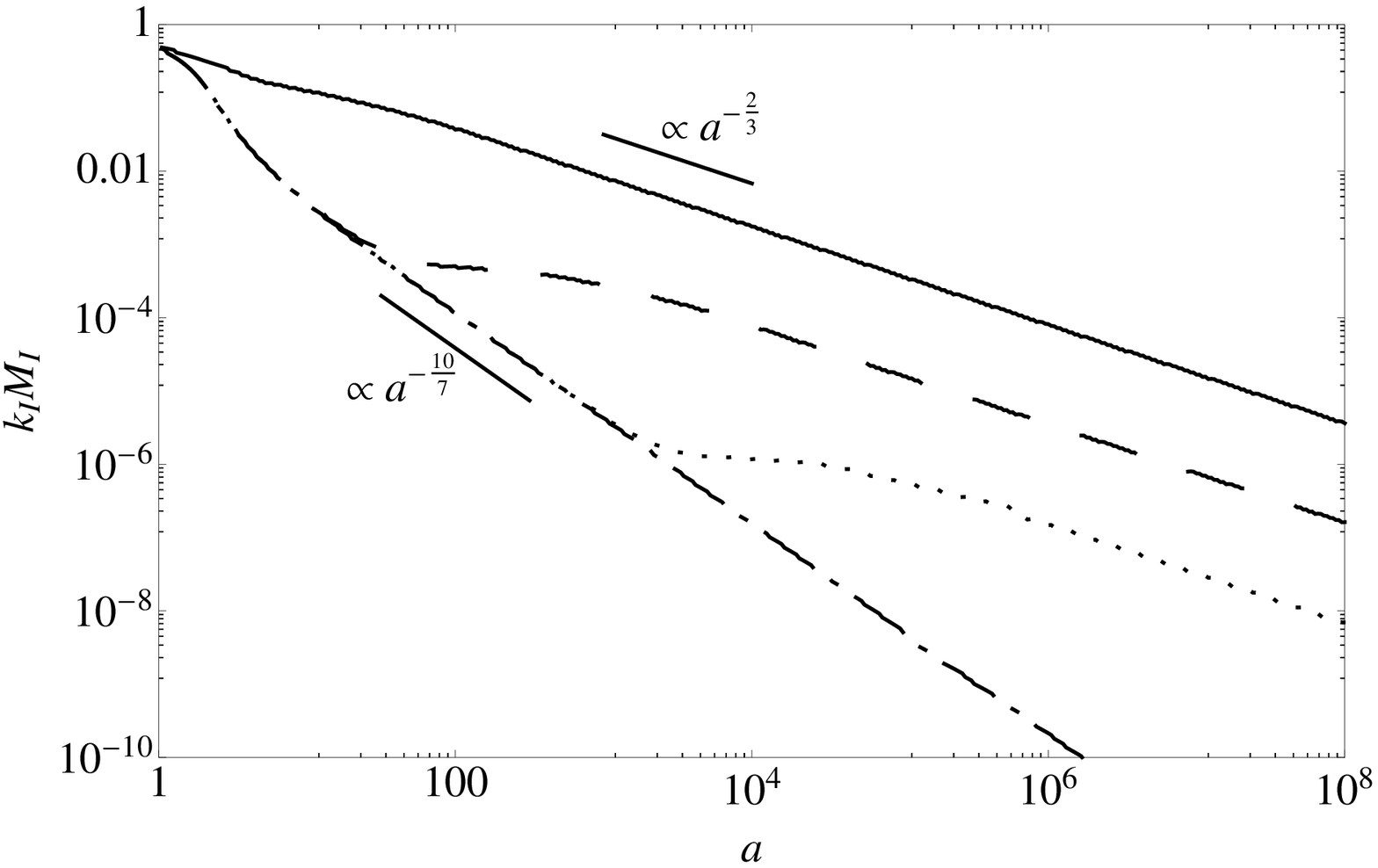}
  \includegraphics[scale=0.467]{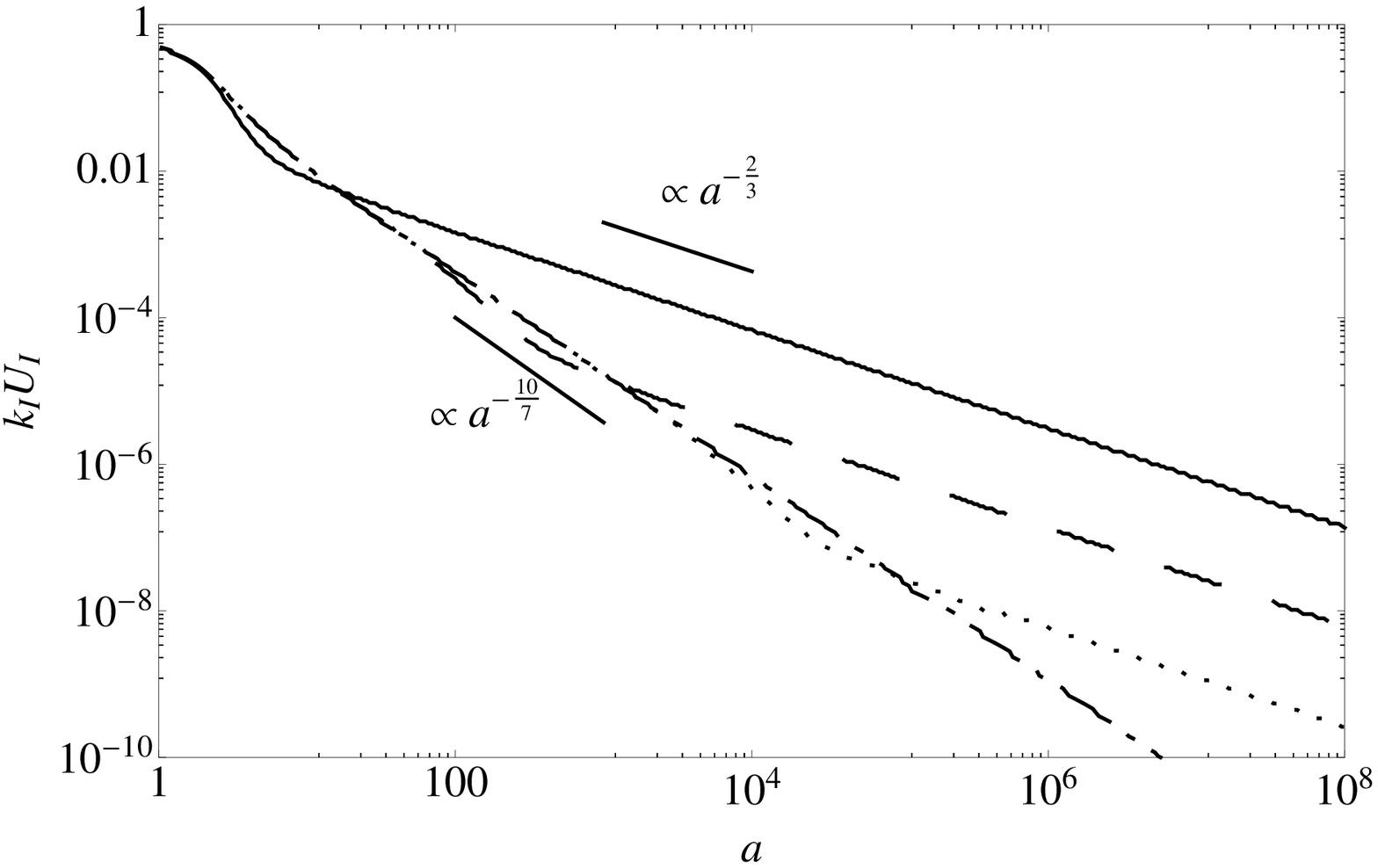}
  \includegraphics[scale=0.467]{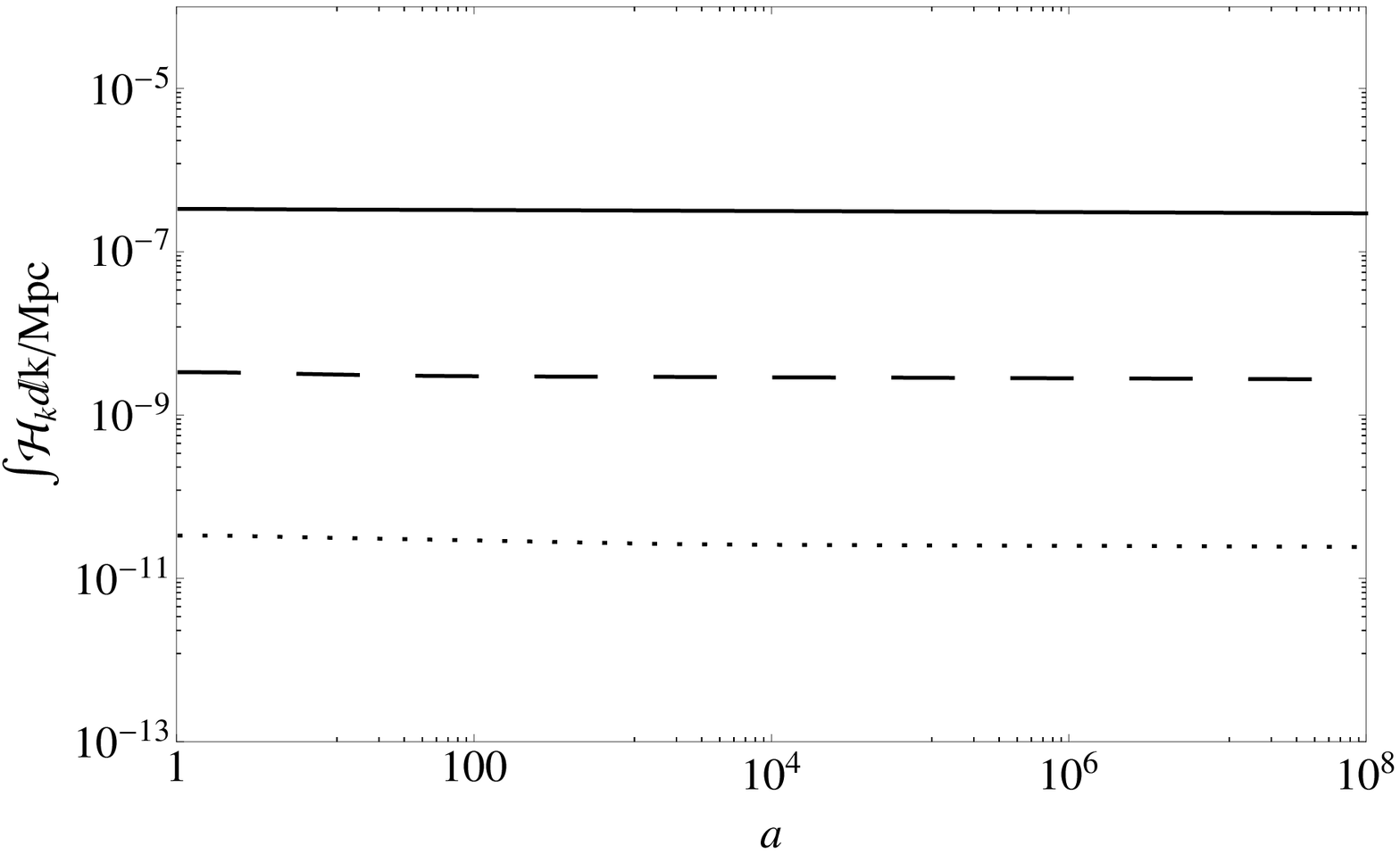}
  \includegraphics[scale=0.6]{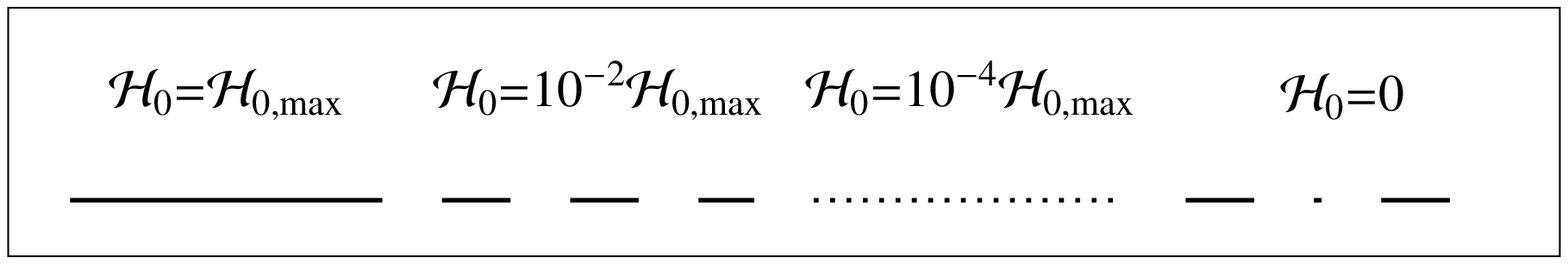}
    \caption{Time evolution of the integral scale ({\it upper left}) and the values at the integral scale of the magnetic spectral energy ({\it upper right}), 
             the kinetic spectral energy ({\it lower left}), and the total normalized magnetic helicity density $h_{B}/\rho = \int \mathcal{H}_{k} {\rm d}k$ 
             ({\it lower right}) for different values of the initial spectral helicity in fractions of its maximal value, $\mathcal{H}_{0,max} \equiv 8 \pi M_{0}/k_{0}$.
             Here we assumed that magnetogenesis has taken place at the QCD epoch.
             }
  \label{fig:kI}
\end{figure*}

\section{Results} \label{sec:Results}
It is now possible to analyze the time development of both the spectral magnetic and kinetic energy as well as the spectral helicity by numerically 
integrating the cosmic version of Eqs.~(\ref{dtMqH})-(\ref{dtHqH}). In comparison to the results without helicity \cite{Saveliev:2012ea}, it may be 
seen from Fig.~\ref{fig:timeevolution} that the absolute value and location of the integral scale (i.e., the scale at which the peak of the 
spectral energy is located) changes dramatically since, due to an inverse cascade, large amounts of energy are transferred to small $q$. On the other 
hand, the scaling of the large-scale tail with the wave number stays the same, i.e., we still have $q M_{q} \sim q^{5}$ and, in addition, we find  
$q \mathcal{H}_{q} \sim q^5$ as well. In contrast to the nonhelical case, however, the slope of the large-scale tail is much less important in 
the maximally helical case. The evolution of magnetic energy density and correlation length is determined solely by the requirement of helicity 
conservation, and not by the tail. In the following we present analytical arguments to explain these results.

\subsection{Large-Scale Magnetic Tail}

As in the case for nonhelical fields~\cite{Saveliev:2012ea}, the nonvanishing term with lowest power in the Taylor expansion in $(q/k)$ of 
Eqs.~(\ref{dtMqH})-(\ref{dtHqH}) is of fourth order for large scales, so that for helical fields as well, we have $B \sim L^{-\frac{5}{2}}$ and $v \sim L^{-\frac{5}{2}}$ 
on large scales. This result is also confirmed by simulation as can be seen in Fig.~\ref{fig:timeevolution}. However, in contrast to nonhelical 
fields, in the fully helical field case, equipartition, i.e., $U_k\simeq M_k$ on all scales, does not hold anymore. Rather it is $U_k\simeq M_k$ only on 
small scales with $U_k << M_k$ on the integral scale, an effect also observed in Ref.~\cite{Banerjee:2004df}.

\begin{figure*}
\centering
  \includegraphics[scale=0.467]{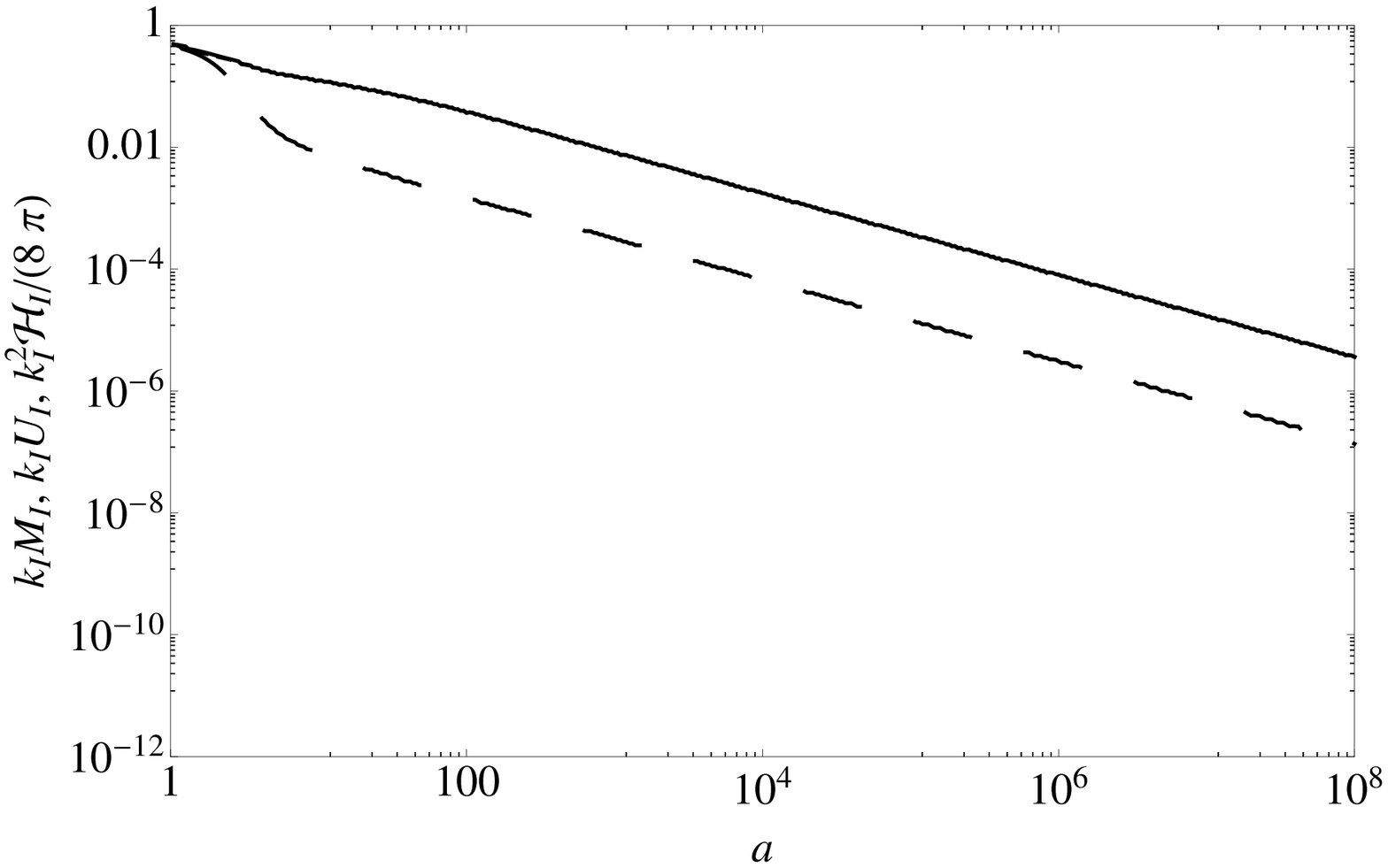}
  \includegraphics[scale=0.467]{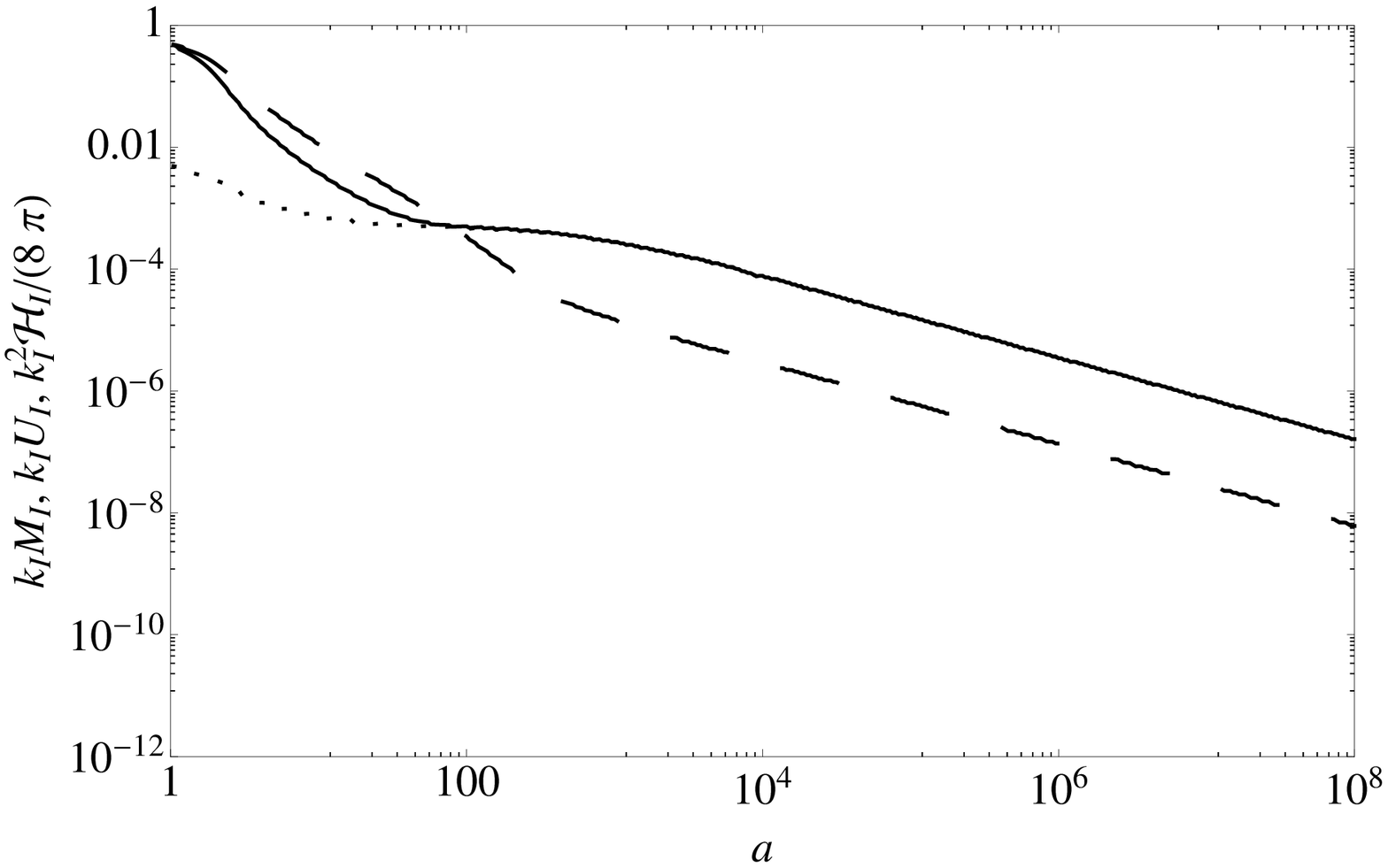}
  \includegraphics[scale=0.467]{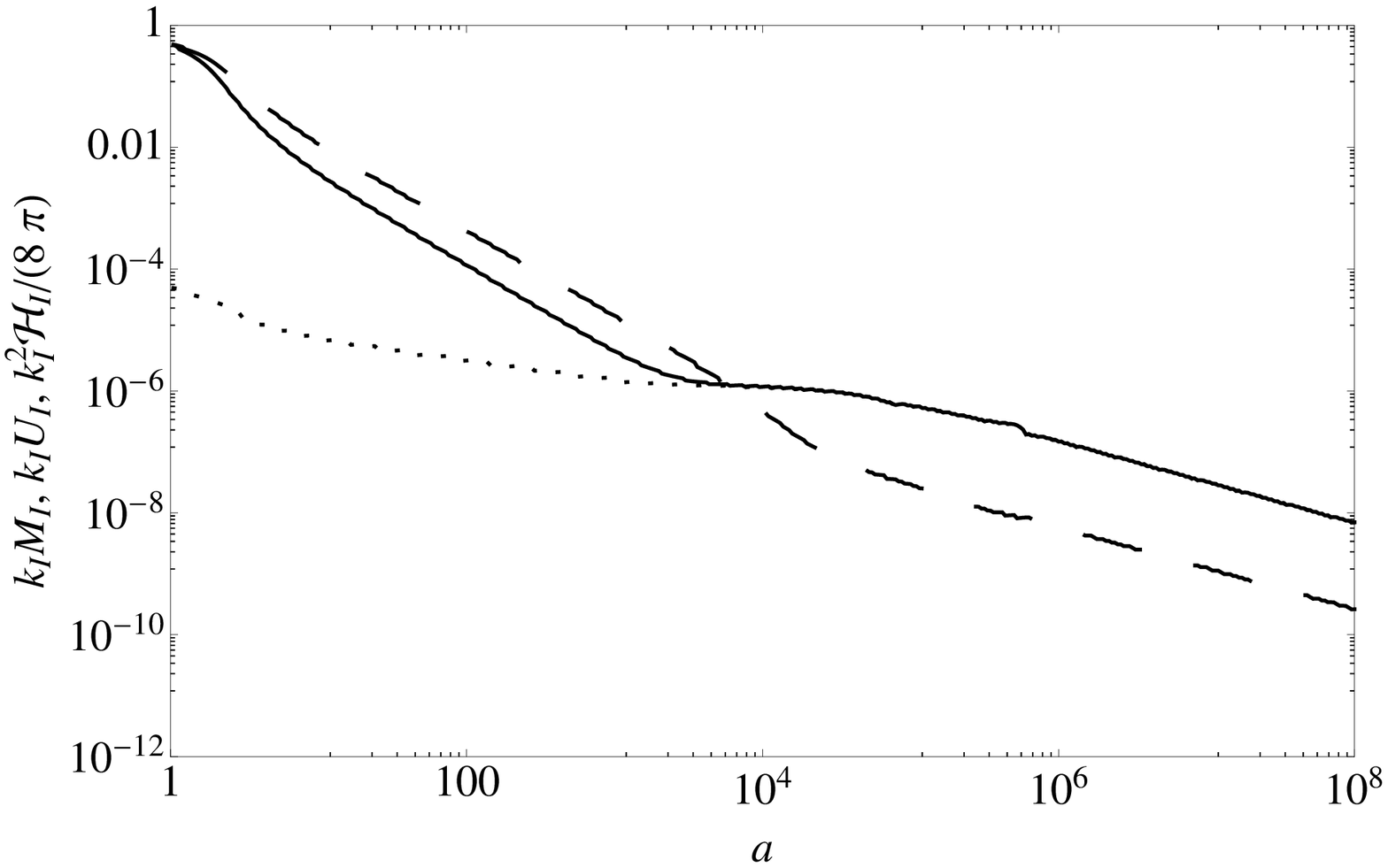} 
  \includegraphics[scale=0.467]{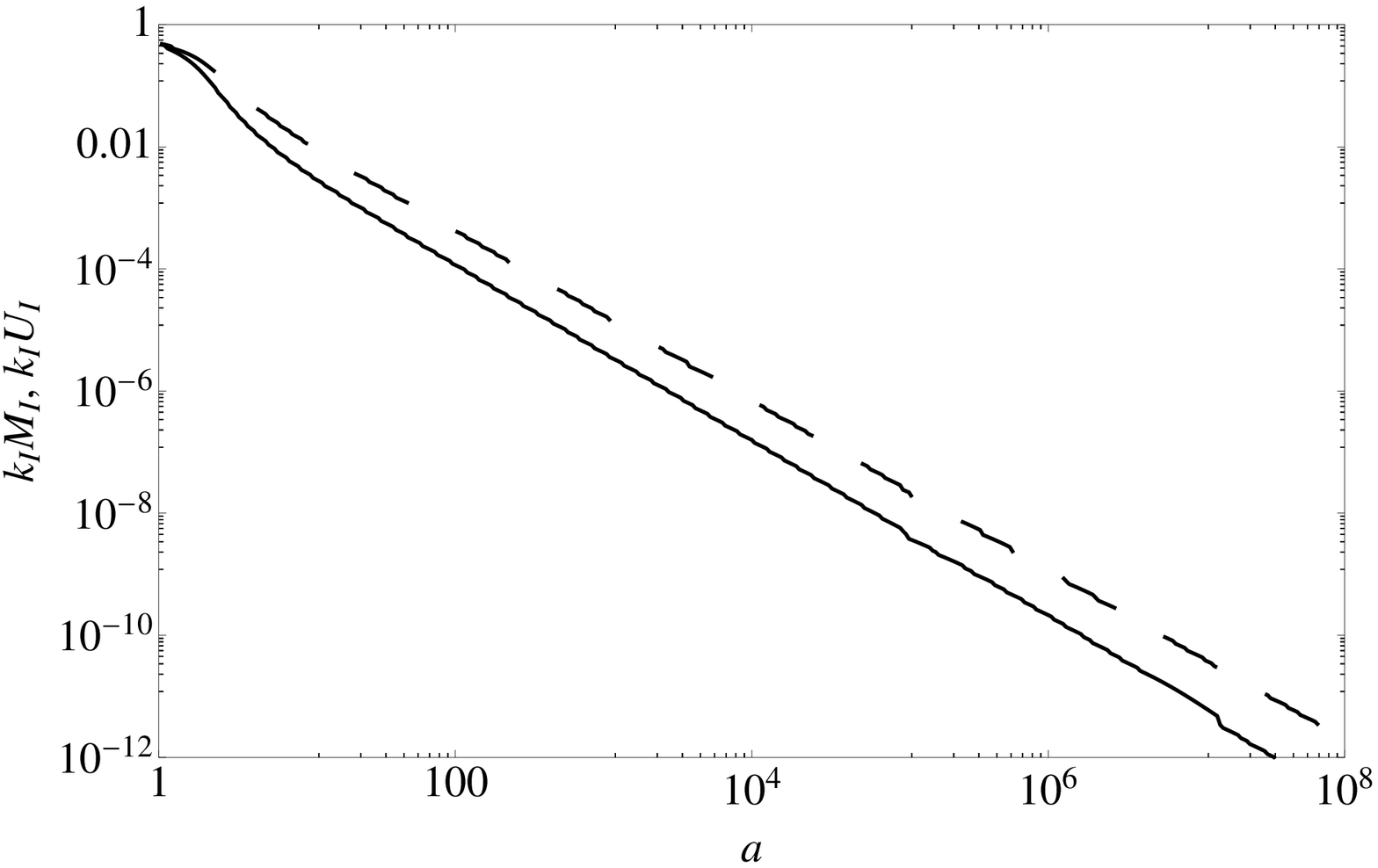}
  \includegraphics[scale=0.6]{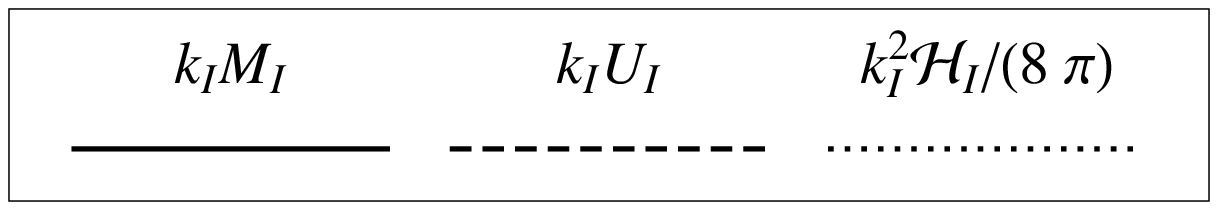}
  \caption{Time evolution of the values of the magnetic and kinetic spectral energies and the helical spectral density at the integral scale, named 
           $M_{I}$, $U_{I}$ and $\mathcal{H}_{I}$, respectively. For convenience these quantities have been multiplied by $k_{I}$ (for $M_{I}$ and 
           $U_{I}$) and by $k_{I}^{2}/(8\pi)$ (for $\mathcal{H}_{I}$). The plots show the situation for different initial values of the helicity 
           $\mathcal{H}_{0}$: maximal helicity, i.e., $\mathcal{H}_{0} = \mathcal{H}_{0,max}$ ({\it upper left}), $\mathcal{H}_{0} = 10^{-2} \mathcal{H}_{0,max}$ 
           ({\it upper right}), $\mathcal{H}_{0} = 10^{-4} \mathcal{H}_{0,max}$ ({\it lower right}) and no helicity, i.e., $\mathcal{H}_{0}=0$ 
           ({\it lower left}). Note that for the case of maximal helicity $k_{I}^{2} \mathcal{H}_{I}/(8\pi)$ has approximately the same value as 
           $k_{I} M_{I}$ and is therefore not visible in the plot. Here we assumed that magnetogenesis has taken place at the QCD epoch.}
  \label{fig:kIEI}
\end{figure*}

\subsection{Evolution of the Integral Scale} \label{sec:kIEvolution}
As it will be shown in the following, the main effect of nonvanishing initial helicity is the dramatic change of the time evolution of the integral 
scale and the corresponding values of the important quantities.

First, we give a short summary of the time evolution of the magnetic spectral energy with vanishing initial helicity. It has been shown \cite{Banerjee:2004df,PhysRevD.83.103005} 
that in this case, if the spectral energy has a slope proportional to $k^{\alpha-1}$ ($\alpha > 1$) for $k<k_{I}$, i.e.,
\begin{equation} \label{qSlope}
\left\langle M_{k} \right\rangle \simeq M_{0} \left(\frac{k}{k_{0}}\right)^{\alpha-1}\,,
\end{equation}
where $M_{0}$ is a normalization constant given by the value of $M_{I}$ at $a_{0}=1$, we have
\begin{equation} \label{kint}
k_{I} \simeq k_{0} a^{-\frac{2}{\alpha+2}}
\end{equation}
for the time-dependent integral scale in the early radiation dominated Universe (i.e., $H=H_0/a^2$), with $k_{0} = k_{I}(a_{0})$ being determined by 
\cite{PhysRevD.83.103005}
\begin{equation} \label{kint0}
k_{0} \simeq \frac{2 \pi H_{0}}{v_{0}} = \left( \frac{4 \pi^{2} H_{0}^{2}}{U_{0}} \right)^{\frac{1}{3}}\,.
\end{equation}
where $U_{0} = U_{I}(a_0)$. Since all the initial magnetic field has been dissipated away for $k>k_I(a)$, the magnetic field left over at scale 
factor $a$ is simply given by Eq.~(\ref{kint}) inserted into Eq.~(\ref{qSlope}), such that we find
\begin{equation} \label{Eint}
k_{I} M_{I} \simeq k_{0} M_{0} a^{-2 \frac{\alpha}{\alpha+2}}\,.
\end{equation}

If, however, the initial helicity is not vanishing, it can be seen from Fig.~\ref{fig:kI} that it only has an influence on the magnetic and kinetic 
spectral energies if at the integral scale it has a value which is close to the maximal spectral helicity value $\mathcal{H}_{I,max}$ given by
\begin{equation} \label{Hmax}
\mathcal{H}_{I,max} = 8 \pi \frac{M_{I}}{k_{I}}\,.
\end{equation}
However, the magnetic spectral energy decays faster than the spectral helicity in the early Universe, and therefore even by starting with a rather small 
initial helicity it is inevitable that after some time it obtains its largest possible value given by Eq.~(\ref{Hmax}). Therefore, there are two successive 
regimes in the time evolution - the first (referred to by ``1'' in the following) where the effect of helicity is negligible and the second (referred to 
by ``2'') where helicity has its maximal value and dominates the time evolution.

One can estimate the time it takes for some initial conditions to evolve to the point where a transition from regime 1 to regime 2 takes place. Here we 
will focus on the magnetic spectral energy as it is the important observable in our case. We are starting with some initial conditions where for 
$a = a_{0} \equiv 1$ we have 
\begin{equation}
k_{I} M_{I} = k_{0} M_{0},\quad \frac{k_{I}^{2} \mathcal{H}_{I}}{8 \pi} = \frac{k_{0}^{2} \mathcal{H}_{0}}{8 \pi} = f_{0} k_{0} M_{0}\, ,
\label{eq:initial} 
\end{equation}
such that $f_{0}=1$ corresponds to maximal initial helicity and $f_{0}=0$ means no initial helicity at all, i.e., $\mathcal{H}_{0} = f_{0} \mathcal{H}_{0,max} $ . 
If we now assume that, as seen in Fig.~\ref{fig:kIEI}, the time evolution for both has a power law dependence on the scale factor $a$, i.e.,
\begin{equation}
k_{I} M_{I} \propto a^{-\mu_{1}},\,k_{I}^{2} \mathcal{H}_{I} \propto a^{-\chi_{1}}\,,
\end{equation}
where, as motivated above, $\mu_{1} > \chi_{1} > 0$, and we can estimate the scale factor $a_{tr}$ of the transition from regime 1 to regime 2 by solving the 
condition $k_{I} M_{I} = k_{I}^2 \mathcal{H}_{I}/(8 \pi)$, i.e.,
\begin{equation}
k_{0} M_{0} a^{-\mu_{1}} = f_{0} k_{0} M_{0} a^{-\chi_{1}}\,,
\end{equation}
for $a$ which gives
\begin{equation}
a_{tr} = f_{0}^{-\frac{1}{\mu_{1}-\chi_{1}}}\,.
\label{eq:atr}
\end{equation}
From this we can, therefore, derive a lower bound on $f_{0}$, such that for all $f_{0}$ larger than this lower bound $f_{min}$, the field has become 
maximally helical by recombination at scale factor $a_{rec}$,
\begin{equation}
f_{min} \simeq a_{rec}^{\chi_{1} - \mu_{1}}\,.
\end{equation}
Here the epoch of recombination is relevant, rather than the present, since, as shown in Ref.~\cite{Banerjee:2004df}, during matter domination after 
the epoch of recombination only negligible further processing of the magnetic fields occurs, provided the magnetic fields are sufficiently strong. 
Taking the values which have been obtained, i.e., $\mu_{1} = \frac{2 \alpha}{\alpha+2} = \frac{10}{7}$ for $\alpha=5$ and $\chi_{1} \simeq 0.32$, 
for $a_{rec} = 10^{8}$ (implying magnetogenesis during the QCD epoch at $T\sim 30\,$MeV)  we therefore get $f_{min} \simeq 7 \times 10^{-10}$. This 
implies that even the smallest initial helicities ultimately have an effect on the time evolution of the fields. Furthermore, also the integral scale 
itself changes its dependence on the scale factor when entering the maximally helical regime 2, i.e., $k_{I} \propto a^{-\kappa_{1}}$ in regime 1, 
where, according to (\ref{kint}), we have $\kappa_{1} = \frac{2}{\alpha+2}$, matching well the numerically found value, and $k_{I} \propto a^{-\kappa_{2}}$ 
in regime 2, where we have numerically obtained $\kappa_{2} \simeq 0.66$ and $\mu_{2} \simeq 0.67$. 

The values $\kappa_2$ and $\mu_{2}$, as well as $\chi_1$, may be deduced unambiguously by the argument of helicity conservation 
\begin{equation} 
h_{B} \simeq L_{I} B_{I}^{2} \simeq \rho k_{I} \mathcal{H}_{I}\simeq h_{B,0}\, ,
\label{eq:helicity}
\end{equation}
where $L_{I} = 2 \pi / k_{I}$ is the integral scale and we have assumed that this scale carries the bulk of the helicity. From Eqs.~(\ref{eq:helicity}) 
and (\ref{kint}) one finds immediately $\chi_1 = - \frac{2}{\alpha+2} \simeq 0.286$ for $\alpha = 5$, to be compared to the numerically found 
$0.3$. The values of $\kappa_{2}$ and $\mu_{2}$ are obtained by considering the turbulent cascading of energy from the integral scale to the 
dissipation scale. This process will decrease $B_{I}$ and will increase $L_{I}$. How fast this can happen is determined by the requirement that an eddy 
turnover can take place on scale $L_{I}$ at epoch with scale factor $a$~\cite{Banerjee:2004df}, in particular
\begin{equation}
\frac{v_{A}^2}{L_{I}^{2} a^{2}} \simeq \frac{B_{I}^{2}}{\rho}\frac{1}{L_{I}^{2} a^{2}} \simeq H\simeq \frac{\rho}{a^{4}}\,.
\end{equation}
It needs to be stressed here that the above is a condition to be formulated with physical quantities as opposed to comoving ones. Since we have dropped
throughout the index ``$c$'', i.e., all quantities in the above equations are comoving, additional powers of the scale factors $a$ enter through 
$L_{I}^{ph}=L_{I} a$ and $\rho^{ph}=\rho/a^{4}$ (note that $B_{I}^{ph} / \sqrt{\rho^{ph}}=B_{I} / \sqrt{\rho}$). Using these equations we find
\begin{equation}
L_{I} \simeq L_{0} a^{2/3}\quad B_I\simeq B_{0} a^{-1/3}\, ,
\end{equation}
for the evolution of comoving integral scale and magnetic field, or equivalently,
\begin{equation}
k_I \simeq k_{0} a^{-2/3}\quad k_{I} M_{I} \simeq k_{0} M_{0} a^{-2/3}
\end{equation}
such that $\kappa_2 = 2/3$ and $\mu_2 = 2/3$, which are in extremely good agreement with the numerical simulations. To summarize the time evolution 
we may write
\begin{equation} \label{kIa}
k_{I}(a) \simeq \begin{cases}&k_{0} a^{-\kappa_{1}},~a \le a_{tr}\\&k_{0} a_{tr}^{\kappa_{2}-\kappa_{1}} a^{-\kappa_{2}},~a > a_{tr}\end{cases}
\end{equation}
with $\kappa_{1} = \frac{2}{\alpha+2}$ and $\kappa_{2} = 2/3$  and
\begin{equation} \label{kIMIa}
k_{I}M_{I}(a) \simeq \begin{cases}&k_{0} M_{0} a^{-\mu_{1}},~a \le a_{tr}\\&k_{0} M_{0} a_{tr}^{\mu_{2}-\mu_{1}} a^{-\mu_{2}},~a > a_{tr}\end{cases}
\end{equation}
with $\mu_{1} = 2 \frac{\alpha}{\alpha+2}$ and $\mu_{2} = 2/3$, where $a_{tr}$ is given by Eq.~(\ref{eq:atr}) with $\mu_1 = 2 \frac{\alpha}{\alpha+2}$
and $\chi_1 = \frac{2}{\alpha + 2}$ for the initial conditions given by Eq.~(\ref{eq:initial}).

\begin{figure*}
\centering
  \includegraphics[scale=0.467]{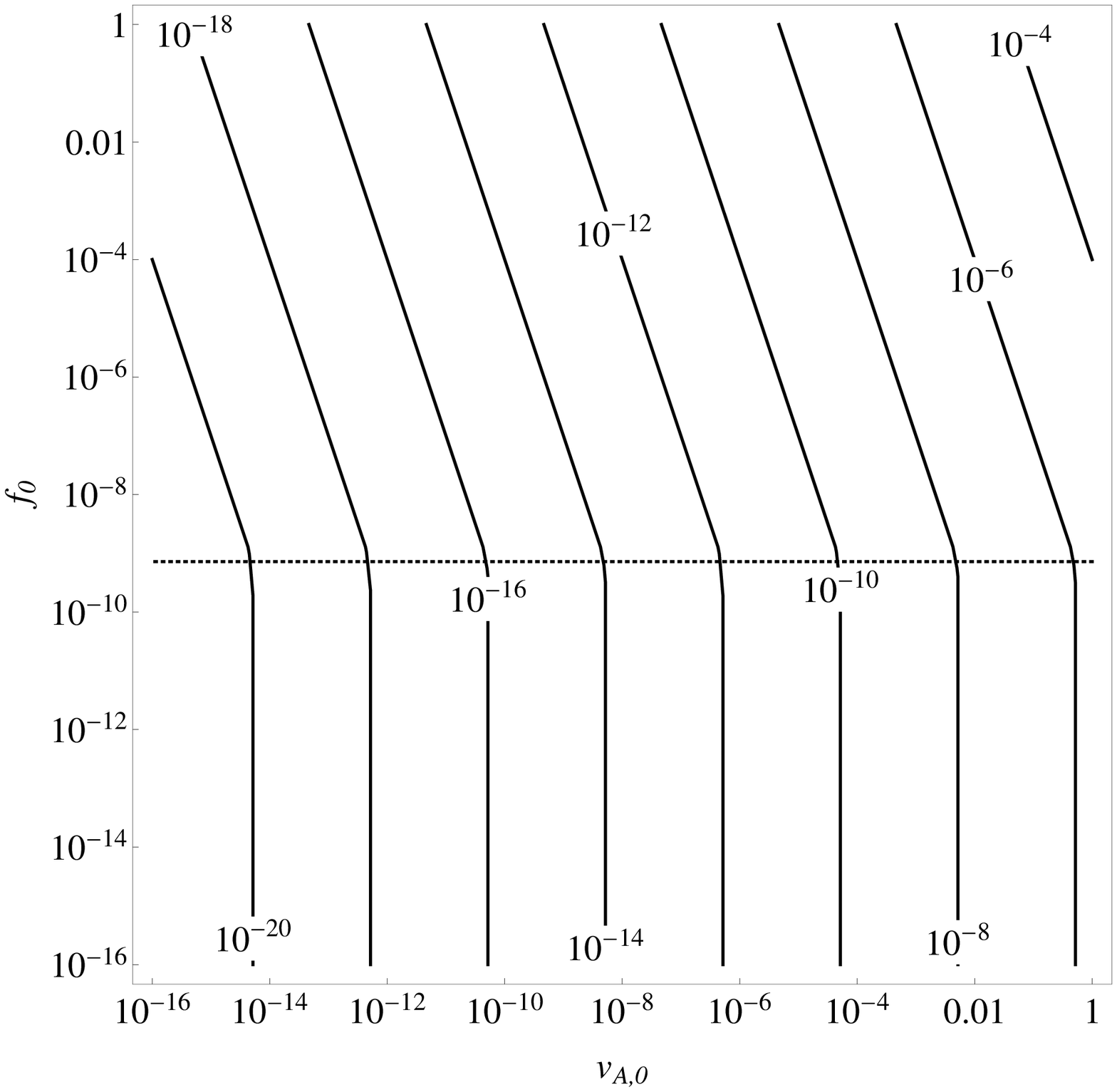}
  \includegraphics[scale=0.467]{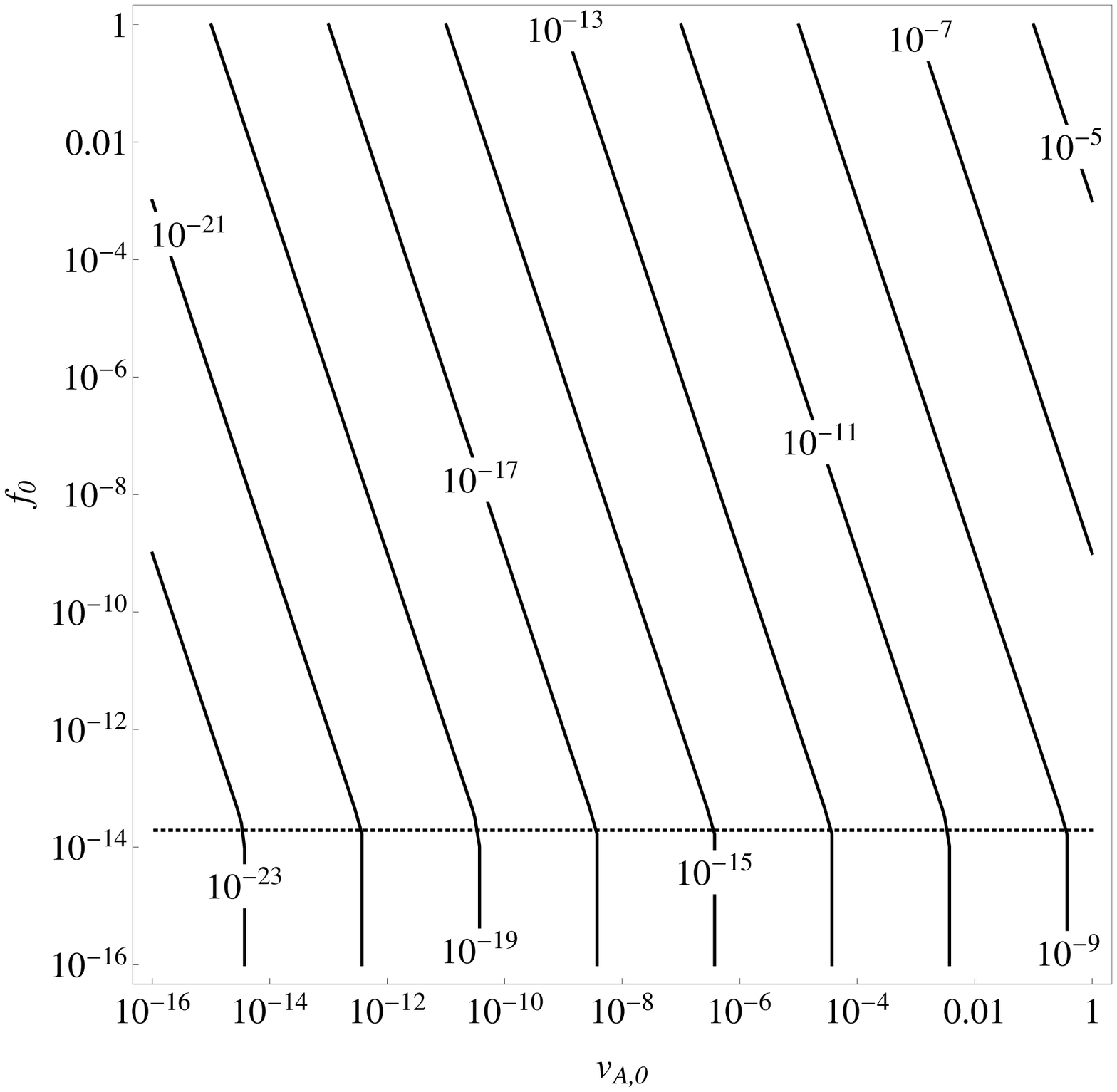} 
  \caption{Present day magnetic field strength having survived dissipation during the evolution in the early Universe as a function of initial 
           Alfv\'en velocity $v_{A,0}$ and helicity $\mathcal{H}_{0} = f_{0} \mathcal{H}_{0,max}$ for magnetogenesis occurring during the QCD epoch 
           ({\it left panel}) and electroweak epoch ({\it right panel}), respectively. Here all initial conditions above the dotted line lead to a 
           maximally helical present day field. Note that $\epsilon_{B} \simeq \rho v_{A}^2/2$. It is stressed that the magnetic field has to fulfill 
           $v_{A} / L \simeq H$ at the magnetogenesis epoch. For different initial conditions, the modified $v_{A,0}$ and $f_{0}$ to be used in the figure may be 
           deduced from Eq.~(\ref{eq:convert}).}
  \label{fig:prediction}
\end{figure*}

\subsection{Enhancement of the Energy Content and the Magnitude of the Magnetic Field}
With these arguments at hand one can now estimate the strength of the magnetic field, which, compared to the case without magnetic helicity, is larger. 
In order to make a reasonable estimate, it is important to take into account the time evolution of both the integral scale $k_{I}$ and the corresponding 
spectral magnetic energy $M_{I}$ in the two different regimes introduced in Sec.~\ref{sec:kIEvolution}, given by Eqs.~(\ref{kIa}) and (\ref{kIMIa}). Using 
the formula for the magnetic field from \cite{Saveliev:2012ea},
\begin{equation}
B(L) = \left(8 \pi k \rho M_{k} \right)^{\frac{1}{2}} = \left(2 k M_{k} \right)^{\frac{1}{2}} B_{0}\,,
\end{equation}
where $B_{0} \simeq (4 \pi \rho)^{\frac{1}{2}}$ is the effective magnetic field for the Alfv\'en velocity $v_{A} \simeq 1$, i.e., the speed of light, 
for the optimistic case of initial equipartition between radiation and magnetic energies. This case would correspond to $B_{0} \simeq 3 \times 10^{-6}\,{\rm G}$. 
If we now acknowledge the integral scale to be the coherence scale of the magnetic field, then for the magnetic field strength at $L_{I} = 2 \pi / k_{I}$ 
we get
\begin{equation}
B(L_{I}) =  \left(2 k_{I} M_{I} \right)^{\frac{1}{2}} B_{0}.
\end{equation}
Using Eq.~(\ref{kIMIa}) we may then finally estimate the magnitude of the present day magnetic field depending on its initial value and the helicity,
\begin{equation} \label{Bfinal}
\frac{B_{\rm today}}{3\times 10^{-6}{\rm Gauss}} \simeq 
\begin{cases}
&v_{A,0} a_{rec}^{-5/7},~f_{0} \le a_{rec}^{-8/7}\\
&v_{A,0} f_{0}^{1/3} a_{rec}^{-1/3},~f_{0} > a_{rec}^{-8/7}\,,
\end{cases}
\end{equation}
where $v_{A,0}$ is the initial Alfv\'en velocity at magnetogenesis, i.e., $v_{A,0}^{2} \simeq 2\epsilon_B/\rho$, and where we have assumed the preferred 
exponent $\alpha = 5$. The surviving magnetic field strength is enhanced in the helical case by $a_{rec}^{8/21} f_{0}^{1/3}$ over that in the nonhelical 
case, which amounts to a factor $2.7\times 10^{3} f_{0}^{1/3}$ and $3.7 \times 10^4 f_{0}^{1/3}$ for magnetogenesis during the QCD and electroweak epoch, 
respectively, highlighting the importance of helicity. The detailed prediction for the present magnetic field strength as a function of the initial 
magnetic field strength and helicity is shown in Fig.~\ref{fig:prediction}, where the QCD and electroweak epochs have been assumed as the epoch of 
magnetogenesis, respectively. In this figure it is assumed that $v_{A} /L\simeq H$ at the magnetogenesis epoch. If this is not the case, the 
appropriate initial conditions to be used in Fig.~\ref{fig:prediction} may be obtained via
\begin{eqnarray} \label{eq:convert}
v_{A,0} \simeq v_{A}^{ini} L_{ini}^{\frac{\alpha}{\alpha+2}} \biggl(\frac{v_{A}^{ini}}{H}\biggr)^{-\frac{\alpha}{\alpha + 2}}\\
f_{0} \simeq f_{ini} L_{ini}^{2 \frac{\alpha - 1}{\alpha+2}} \biggl(\frac{v_{A}^{ini}}{H}\biggr)^{-2 \frac{\alpha - 1}{\alpha + 2}}
\nonumber
\end{eqnarray}
unless $f_{0} > 1$ or $v_{A,0} > v_A^{ini}$. Here $v_A^{ini}$, $L_{ini}$, and $f_{ini}$ quantify the initial total magnetic field strength, coherence length 
and helicity. Note that $\alpha = 5$ should be employed in Eq.~(\ref{eq:convert}), as such a large scale magnetic field tail will quickly develop 
independent of the initial conditions~\cite{Saveliev:2012ea}. 

\section{Conclusions} \label{sec:Conclusions}

In this paper we have extended our recent derivation~\cite{Saveliev:2012ea} of the evolution of the spectral magnetic and kinetic energy densities in 
homogeneous and isotropic magnetohydrodynamics by a closure theory to include helicity. The resulting ordinary differential equations were subsequently 
numerically integrated to predict the evolution of helical cosmic magnetic fields from the early Universe to the present. We find that the energy 
content and coherence length of submaximally helical magnetic fields essentially evolve as those of magnetic fields in the completely nonhelical 
case. Due to the long cosmic expansion from the early Universe to the present, however, even magnetic fields with the smallest initial helicities 
ultimately become maximally helical before the present epoch, thereafter decaying slower than nonhelical magnetic fields due to an inverse cascade. 
Our simulations agree well with all prior analytic estimates on the evolution of helical magnetic fields. 

\begin{acknowledgments}
This work was supported by the Deutsche Forschungsgemeinschaft through the collaborative research centre SFB 676, by the Helmholtz Alliance for
Astroparticle Phyics (HAP) funded by the Initiative and Networking Fund of the Helmholtz Association, and by the State of Hamburg through the
Collaborative Research program ``Connecting Particles with the Cosmos''.
\end{acknowledgments}

\appendix
\section{Derivation of the Master Equations (\ref{dtMqH}), (\ref{dtUqH}) and (\ref{dtHqH})} \label{app:calc}
The derivation of these equations follows exactly the same lines as the one presented in Appendix A of Ref.~\cite{Saveliev:2012ea} for nonhelical 
magnetic fields. In the more general case presented here, magnetic field correlators are assumed to be
\begin{widetext}
\begin{eqnarray} \label{corr1}
\left\langle \hat{B}_{a}(\mathbf{k},t) \hat{B}_{b}(\mathbf{k'},t')^{*} \right\rangle &\simeq& C_{1} \delta_{\mathbf{k}\mathbf{k'}} \delta_{tt'} M_{k}\left( \delta_{ab} - \frac{k_{a} k_{b}}{k^{2}} \right) + C_{3} \delta_{\mathbf{k}\mathbf{k'}} \delta_{tt'} i \epsilon_{abc} k_{c} \mathcal{H}_{k}\, , \label{ensrules1}
\end{eqnarray}
with the first and second part of the correlator describing the nonhelical and helical components of the magnetic field, respectively.

\section{Alternative form of the Master Equations} \label{app:alternative}
We give here an alternative formulation of the master equations (\ref{dtMqH})-(\ref{dtHqH}) in terms of $\mathbf{k_{1}} = \mathbf{q} - \mathbf{k}$ 
which is more suitable for numerical integration:
\begin{equation} \label{dtMqfinalKJk1}
\begin{split} 
&\left\langle \partial_{t} M_{q} \right\rangle = \int_{0}^{\infty} {\rm d}k \Bigg( \Delta t \Bigg\{ -\frac{2}{3} q^2 \left\langle M_{q} \right\rangle \left\langle U_{k} \right\rangle - \frac{4}{3} q^{2} \left\langle M_{q} \right\rangle \left\langle M_{k} \right\rangle + \frac{1}{3} \frac{1}{(4 \pi)^{2}} q^{2} k^{2} \left\langle \mathcal{H}_{q} \right\rangle \left\langle \mathcal{H}_{k} \right\rangle \\
&+ \int_{|q-k|}^{q+k} { \rm d}k_{1} \left[ \left( - \frac{q^{7}}{16 k^{3} k_{1}^{3}} + \frac{q^{5}}{16 k^{3} k_{1}} + \frac{q^{5}}{16 k k_{1}^{3}} + \frac{ q^{3} k}{16 k_{1}^{3}} + \frac{3 q^{3}}{8 k k_{1}} + \frac{q^{3} k_{1}}{16 k^{3}} - \frac{q k^{3}}{16 k_{1}^{3}} + \frac{q k}{16 k_{1}} + \frac{q k_{1}}{16 k} - \frac{q k_{1}^{3}}{16 k^{3}} \right) \left\langle M_{k} \right\rangle \left\langle U_{k_{1}} \right\rangle \right] \Bigg\} \Bigg)
\end{split}
\end{equation}
and
\begin{equation} \label{dtUqfinalKJk1}
\begin{split}
&\left\langle \partial_{t} U_{q} \right\rangle = \int_{0}^{\infty} {\rm d}k \Bigg(  \Delta t \Bigg\{ - \frac{2}{3} q^{2} \left\langle M_{k} \right\rangle \left\langle U_{q} \right\rangle - \frac{2}{3} q^{2} \left\langle U_{q} \right\rangle \left\langle U_{k} \right\rangle\\ 
&+ \int_{|q-k|}^{q+k} {\rm d}k_{1} \Bigg[ \left( - \frac{q^{5}}{16 k k_{1}^{3}} + \frac{q^{3} k}{8 k_{1}^{3}} + \frac{3 q^{3}}{8 k k_{1}} - \frac{q k^{3}}{16 k_{1}^{3}} + \frac{3 q k}{8 k_{1}} - \frac{5 q k_{1}}{16 k} \right) \left\langle M_{k} \right\rangle \left\langle M_{k_{1}} \right\rangle \\
&+ \left( \frac{q^{7}}{32 k^{3} k_{1}^{3}} - \frac{7 q^{5}}{32 k k_{1}^{3}} - \frac{3 q^{5}}{32 k^{3} k_{1}} + \frac{11 q^{3} k}{32 k_{1}^{3}}+ \frac{5 q^{3}}{16 k k_{1}} + \frac{3 q^{3} k_{1}}{32 k^{3}} - \frac{5 q k^{3}}{32 k_{1}^{3}} + \frac{9 q k}{32 k_{1}} - \frac{3 q k_{1}}{32 k} - \frac{q k_{1}^{3}}{32 k^{3}} \right) \left\langle U_{k} \right\rangle \left\langle U_{k_{1}} \right\rangle \\
&+ \frac{1}{(8 \pi)^{2}} \left( \frac{q^{5}}{16 k k_{1}} - \frac{3 q^{3} k}{8 k_{1}} - \frac{q^{3} k_{1}}{8 k} + \frac{5 q k^{3}}{16 k_{1}} - \frac{3 q k k_{1}}{8} + \frac{q k_{1}^{3}}{16 k} \right) \left\langle \mathcal{H}_{k} \right\rangle \left\langle \mathcal{H}_{k_{1}} \right\rangle \Bigg] \Bigg\} \Bigg)\,,
\end{split}
\end{equation}
as well as
\begin{equation} \label{dtHqfinalKJk1}
\begin{split}
\langle \partial_{t} \mathcal{H}_{q} \rangle &= \int_{0}^{\infty} {\rm d}k \Bigg( \Delta t \Bigg\{ \frac{4}{3} k^{2} \langle M_{q} \rangle \langle \mathcal{H}_{k} \rangle - \frac{4}{3} q^{2} \langle M_{k} \rangle \langle \mathcal{H}_{q} \rangle - \frac{2}{3} q^{2} \langle U_{k} \rangle \langle \mathcal{H}_{q} \rangle \\
&+ \int_{|q-k|}^{q+k} { \rm d}k_{1} \Bigg[ \left( - \frac{q^{5}}{8 k k_{1}^{3}} + \frac{q^{3} k}{4 k_{1}^{3}} + \frac{q^{3}}{4 k k_{1}} - \frac{q k^{3}}{8 k_{1}^{3}} + \frac{q k}{4 k_{1}} - \frac{q k_{1}}{8 k} \right) \langle U_{k_{1}} \rangle \langle \mathcal{H}_{k} \rangle \Bigg] \Bigg\} \Bigg)\,.
\end{split}
\end{equation}
We note here that for some terms a substantially shorter expression than that given in Appendix B of Ref.~\cite{Saveliev:2012ea} has been found by 
realizing that $\int_0^{\infty}{\rm d}k\int_{|q-k|}^{q+k} { \rm d}k_{1}$ is equivalent to $\int_0^{\infty}{\rm d}k_{1}\int_{|q-k|}^{q+k} { \rm d}k$. 
\end{widetext}

\end{document}